# Lattice dynamical origin of peak thermoelectric performance in AgPb$_m$SbTe$_{2+m}$ observed by inelastic neutron scattering

M. E. Manley[1], S. Shapiro[2], Q. Li[2], A. Llobet[3], M. E. Hagen[4]

[1]*Lawrence Livermore National Laboratory, Livermore, California 94550, USA*
[2]*Brookhaven National Laboratory, Upton, New York 11973, USA*
[3]*Los Alamos National Laboratory, Los Alamos, New Mexico 87545, USA*
[4]*Oak Ridge National Laboratory, Oak Ridge, Tennessee 37830, USA*

Phonon densities of states (DOS) for the high performing thermoelectric material, AgPb$_m$SbTe$_{2+m}$ (LAST-$m$, $m$ = 16, 18, and 20), were extracted from time-of-flight inelastic neutron scattering measurements. The phonon DOS of LAST-18 differs remarkably from LAST-16 and LAST-20 by exhibiting a dramatic broadening of its acoustic modes that increases on heating. This broadening coincides with a minimum in the thermal conductivity, a maximum in the electrical conductivity and Seebeck coefficient, and a related peak in thermoelectric performance. We argue that the anomalous broadening originates with scattering enhanced by modifications to Te-Ag(Sb) bonds caused by their resonant electronic states falling near the Fermi energy for $m$ = 18.

## I. INTRODUCTION

Thermoelectric materials are used in devices designed to convert heat into electrical energy or electrical energy into cooling power and have become the subject of intense interest in recent years.[1-7] A material's figure-of-merit (FOM) used to judge its suitability as a thermoelectric is: $ZT=\alpha^2\sigma T/\kappa$, where $\alpha$ is the Seeback coefficient, $\sigma$ is the electrical conductivity, and $\kappa$ is the thermal conductivity. Performance is improved by minimizing thermal conductivity while increasing the power factor $\alpha^2\sigma$. This can be achieved in part by reducing the lattice component of the thermal conductivity, which is proportional to the phonon mean free path and group velocity. The group velocity is typically highest in the acoustic branches, as they will often have the steepest dispersion. Hence, it is most desirable to reduce the mean free path of the acoustic phonons. The mean free path is reduced through phonon scattering processes and it has been argued that such scattering processes can be enhanced by introducing inhomogeneity or heavy "rattler" atoms that form Einstein-like modes across the acoustic branches.[8] However, recent inelastic neutron scattering measurements on skutterudites have shown that even for loosely caged heavy atoms spatial correlations[9] and dispersion[10] occurs in the low lying modes, showing that they are not true "rattlers". In an alternate mechanism, elastic stiffness mismatch in repeated nanoscale structures, as demonstrated for multilayered materials,[6] can also inhibit the thermal conductivity.

A three-dimensional analog of the elastic mismatch mechanism at the nanoscale, as demonstrated for multilayered materials,[6] may also be playing a role in the high thermoelectric performance of AgPb$_m$SbTe$_{2+m}$ (called LAST-$m$).[11-13] The thermal conductivity is sensitive to $m$ and LAST-18 has the highest FOM of known bulk materials, 2.2 at 800K.[11] Detailed structural studies[14,15] have shown that this material does not behave as a classical solid solution but has a range of structural features extending from atomic ordering to nanostructure inhomogeneities. These nanostructure features could be enhancing the phonon scattering and thereby suppressing the thermal conductivity. Indeed, recent first-principles calculations and measurements show that LAST-$m$ is expected to exist as $AgSbTe_2$ nanoscale precipitates in a PbTe matrix,[12,15] and further that the stiffness of phonon density of states (DOS) of the precipitates and matrix are mismatched in the acoustic energy range.[12] Barabash *et al.*[12] suggest three contributing factors to the low thermal conductivity: (i) roughness of the precipitate-matrix interface, (ii) mismatch of the acoustic phonon frequencies between the precipitates and the matrix, and (iii) mismatch of the ionic character of the vibrational modes, causing bending and scattering of acoustic waves.[12] The precipitate-matrix interface was recently addressed in detail using high resolution transmission electron microscopy supported by calculations.[15] However, recent transport measurements by Dow *et al.*[16] show an anomalously sharp minimum in the thermal conductivity of LAST-$m$ at $m = 18$ despite the expected gradual development of the nano-precipitates,[12] suggesting that the existence precipitates alone is not the entire story. In this Letter, we address the lattice dynamical origin of the minimum thermal conductivity using neutron scattering measurements of the phonon DOS of LAST-$m$ with different values of $m$, and compare results to the calculated phonon DOS of Barabash *et al.*[12] and transport properties. While differences between the phonon DOS of LAST-16 and LAST-20 are small and partially understood in terms of small additive changes, LAST-18 stands out by exhibiting a dramatic broadening of its acoustic phonons. The broadening coincides with a sharp minimum in the thermal conductivity, maximum in the electrical conductivity and Seebeck coefficient, and a related maximum in the thermoelectric FOM for LAST-18.[11,16]

## II. EXPERIMENTAL

All samples were polycrystalline and formed by solid state reaction with intermediate ball milling. Samples of LAST-16 and LAST-20 were used directly as powders while LAST-18 was a powder pressed into two pellets prior to measurement. Neutron scattering spectra were obtained on the LANSCE-PHAROS time-of-flight chopper spectrometer located at Los Alamos National Laboratory. The incident neutron energy was 25 meV or 50 meV and measurements were performed at both 7 K and room temperature. The raw data were corrected for sample environment background, detector efficiency, and the $k_i/k_f$ phase space factor. The incoherent multiphonon scattering for each spectrum was determined iteratively to all orders[17] and subtracted from the data.

## III. RESULTS & DISCUSSION

Figure 1 shows the extracted phonon DOS for all three compositions along with the first-principles calculated phonon DOS of PbTe and AgSbTe$_2$ from Barabash et al.[12]. The LAST-16 and LAST-20 phonon DOS show only small differences between each other, Fig. 1a. The positions of the main features are fairly close to those predicted for PbTe but the distribution is different. The measured phonon DOS, Fig. 1a, shows its strongest feature at the lower energy end of the spectra (around 5 meV) while the calculated phonon DOS of the dominant PbTe phase, Fig. 1b, has its strongest feature at the higher energies (around 12 meV). This difference cannot be accounted for by the neutron-cross-section weighting of the phonon DOS. The cross section weighting scales as the neutron cross section divided by mass[18] and is taken into account in calculating the curve in Fig. 1b labeled "Neutron wt. DOS". Apparently, the theory over estimates the energy difference (spread) between the shear and longitudinal modes in the acoustic branches while under estimating the difference (spread) for the optic modes. Some phonon lifetime broadening may also play a role in the differences in the measured and calculated optic modes.

Some of the differences between the measured LAST-16 and LAST-20 can be understood simply by adding the calculated phonon DOS from the AgSbTe$_2$ nano-precipitates (or short range ordered regions[14]) to the calculated phonon DOS of the PbTe matrix. For both cases the PbTe matrix dominates, but the nano-precipitate concentration increases with decreasing $m$. For $m$ = 16 up to 0.125 of the atoms can be in the precipitates relative to the matrix. Figure 1d shows the pure PbTe phonon DOS compared with the phonon DOS composed of PbTe and 0.125 AgSbTe$_2$ (each part normalized to 1 to start). Several of the apparent shifts in the phonon DOS, indicated with arrows in Fig. 1d, are consistent with the differences observed experimentally with decreasing $m$ (increasing nano-precipitate atoms) as indicated by arrows in Fig. 1a; in particular, the small stiffening of the phonons in the ~5 meV feature and the bump near the cutoff energy near 14.5 meV. On the other hand, some changes are inconsistent or at least too large to be accounted for this way; including the small loss of modes around 8 meV, which is inconsistent, and the enhancement at 6 meV which is too large. These effects most likely result from broadening associated with phonon scattering, which can take effect in the dominant PbTe matrix and develop in a non-additive way with increasing $m$ (scattering is a nonlinear effect). There are also expected to be corrections to the nano-precipitate phonons for rough interfaces and coherency strains,[15] but these should only perturb the states that are added in proportion to concentration.

The importance of broadening is most evident in the intermediate LAST-18 phonon DOS, Fig. 1e. At 300 K the sharp feature at 5 meV that is dominant in both LAST-16 and LAST-20 is essentially absent in LAST-18. Cooling to 7 K suppresses the thermal broadening enough to recover the feature in LAST-18 (bottom panel Fig. 1e), but even at 7 K the feature is still broader than it is in either LAST-16 or LAST-20 at 300 K. Higher resolution measurements on LAST-18 using 25 meV neutrons produced similar results as the 50 meV neutron data shown in Fig. 1e.

Interestingly, Dow *et al.*[16] showed that many of the transport properties of LAST-18 are also distinctly different from a whole series of LAST-*m* compounds, including *m* = 12, 16, 20, 22, and 26. Of particular relevance is the thermal conductivity which, when plotted as a function of composition (Fig. 2), shows a sharp minimum at *m* = 18. The occurrence of a minimum in the thermal conductivity coincident with considerable phonon broadening in the acoustic modes (see insets in Fig. 2) is not surprising since both are related to an enhanced rate of phonon scattering. Enhanced phonon scattering produces lifetime broadening in the phonon DOS and implies shorter mean free paths for the phonons, which reduces the lattice thermal conductivity. What is surprising, however, is the occurrence of this suppressed thermal conductivity and enhanced phonon scattering rate at just one composition. The enhanced phonon scattering in LAST-18 also likely extends over a wide range of the acoustic phonon branches. The washing out of the peak in the phonon DOS indicates that the broadening extends to the dispersionless part of the modes near the zone boundary (ZB). These short-wavelength modes, however, have small group velocities and therefore contribute little to the thermal conductivity. The large suppression of the thermal conductivity, therefore, suggests that the enhanced scattering rate also extends to the longer-wavelength steeply dispersing parts of the acoustic modes. Hence, there appears to be something occurring at or near *m* = 18 that leads to enhanced scattering rates in the acoustic modes at both short and long wavelengths. Only inelastic neutron or x-ray scattering experiments on single crystals can observe this directly, but this would be challenging given the nano-precipitates.

The extensive phonon broadening in LAST-18 is difficult to explain by phonon scattering at precipitates alone. First, the short-wavelength ZB modes, which are broadened, exhibit many wavelength oscillations in the space between precipitates and thus should not be appreciably broadened by scattering at this length scale. Second, precipitates are expected to exist for all three compositions[12] yet only the LAST-18 exhibits the severe broadening. This last reason points to an accidental condition. Such an accident may originate with the resonant states in the electronic structure that were demonstrated using *ab initio* calculations for LAST-*m* by Bilc *et al.*[4] and for a wide range of related systems by Hoang *et al.*[19] The resonant states appear near the top of the valence and bottom of the conduction bands of bulk PbTe when Ag and Sb replace Pb, and this can be understood in terms of modified Te-Ag(Sb) bonds.[4] A relationship between sharp features near the Fermi level and phonon softening has recently been demonstrated.[20] In the present case, the locally modified bonds would be expected to enhance phonon scattering owing to the mismatched character of the modes.[12] Other expected manifestations of the resonant states near the Fermi energy include an enhancement of the electrical conductivity since they add electronic states near the Fermi energy[21] and an enhancement of the Seebeck coefficient since it increases the energy derivative at the Fermi energy. Consistent with this notion, both the electrical conductivity and the Seebeck coefficient are especially high in LAST-18 relative to other compositions measured, including $m = 12, 16, 20, 22,$ and $26$.[16] Hence, one explanation for this outlier behavior might be that these resonant states happen to be especially close to the Fermi energy in LAST-18, causing a high degree of phonon scattering (suppressing the thermal conductivity) due to electron-phonon interactions and simultaneously enhancing the electrical conductivity, resulting in the peak in the thermoelectric FOM at $m = 18$.

## IV. CONCLUSIONS

While more work is needed to fully establish the connection between the atomic arrangements at the nanoscale and the strong phonon scattering rate (and suppressed lattice thermal conductivity) in LAST-18, the present results show that the effect on the phonons, as with the transport properties,[16] is not something that develops in a gradual monotonic way with compositional changes, but rather is punctuated in a narrow composition range. This is scientifically interesting because it points to a resonance in the phonon scattering mechanism that may be related to resonant features in the electronic structure.[4] It is well known that resonant structure in the electronic DOS at Fermi level is of particular importance to thermoelectric performance because a delta-shaped transport distribution has long been considered the most desirable to maximize the FOM.[22] A grand challenge in practical thermoelectric research is to use crystal and chemical design to tune the resonant peak to optimize the FOM. This study suggests that there may exist an even more optimum composition (or nanostructure) in the vicinity of $m = 18$ with an even higher FOM.

**ACKNOWLEDGEMENTS**

This work was performed under the auspices of the U.S. Department of Energy by Lawrence Livermore National Laboratory under Contract No. DE-AC52–07NA27344. The work at Brookhaven National Laboratory was supported by the U.S. Department of Energy, Basic

**Figure Captions:**

FIG. 1. Phonon density of states (DOS) of LAST-$m$ measured for $m$ = 16, 18, 20 along with calculated curves after Ref. [12]. (a) Measured phonon DOS of LAST-16 and LAST-20 at 300 K with 50 meV incident energy neutrons. (b) Calculated curves for PbTe after Ref. [12]. The solid black line is the total phonon DOS. The Pb and Te indicate partial DOS by atom and the "Neutron wt. DOS" was calculated accounting for the neutron weighting (cross section divided by mass for each atomic partial DOS). (c) The total phonon DOS of AgSbTe$_2$ after Ref. [12]. (d) The calculated phonon DOS of PbTe along with a DOS composed of PbTe plus 0.125 AgSbTe$_2$. (e) Measured phonon DOS of LAST-18 at 7 K and 300 K with 50 meV incident energy neutrons. The 300 K data is offset for clarity.

FIG 2. Thermal conductivity of LAST-$m$ measured as a function of $m$ at various temperatures as measured by Dow *et al.* [16]. The inset images reveal the relationship between the dip in the thermal conductivity and the broadening of the low energy phonons for $m$ = 18.

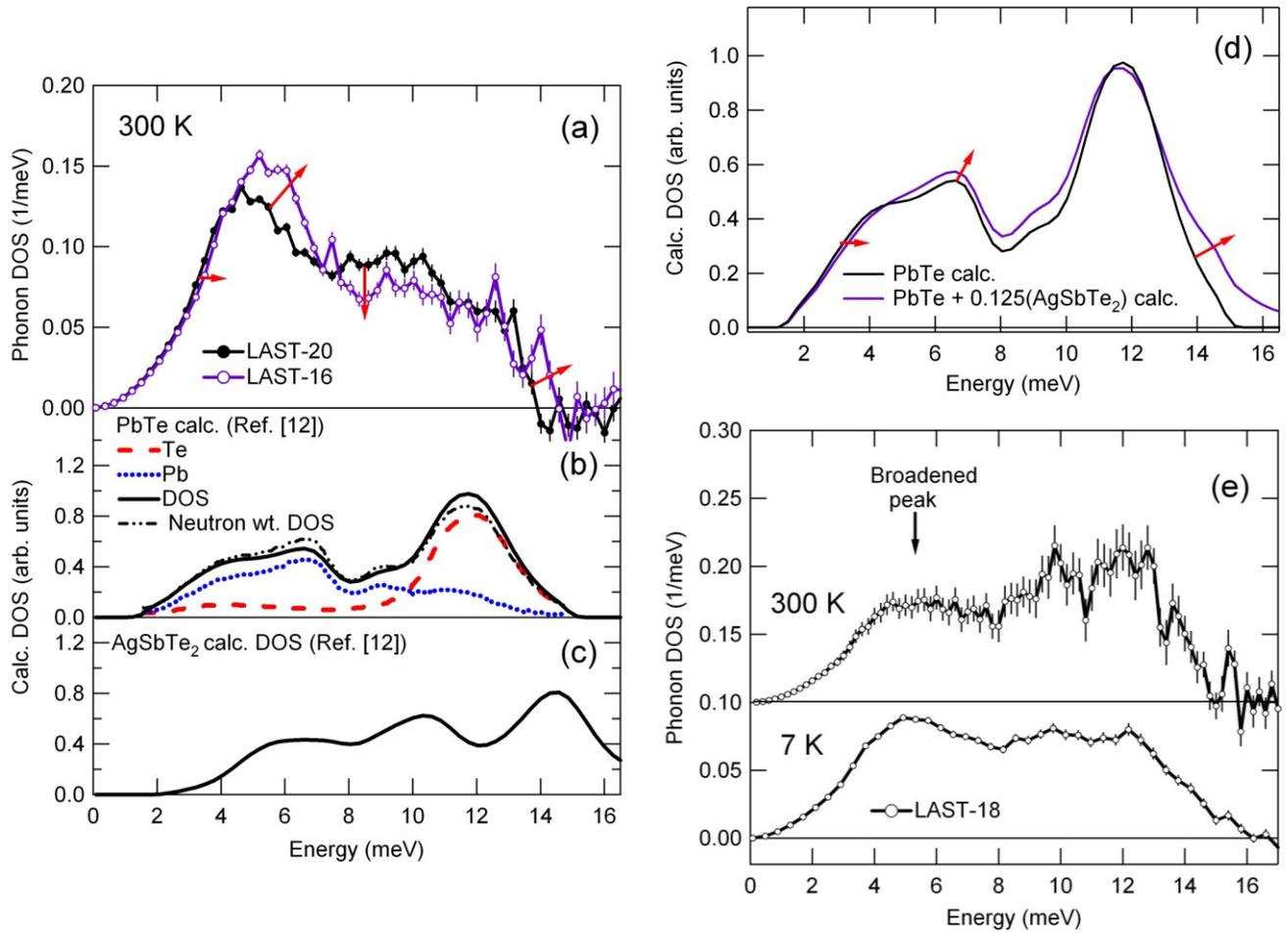

FIG. 1

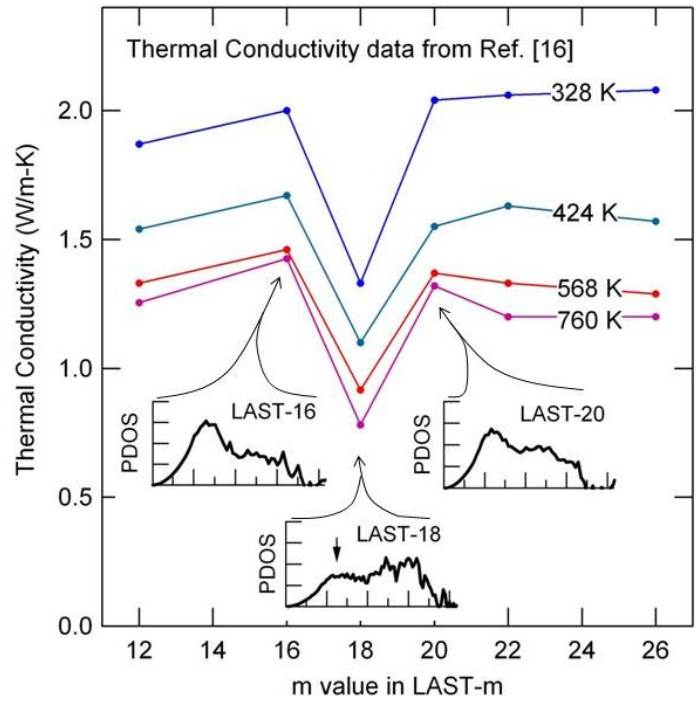

FIG. 2